\documentclass[aps,prd,
preprintnumbers,nofootinbib,superscriptaddress]{revtex4}
\usepackage[dvips]{graphicx}
\usepackage{caption}
\usepackage{bm,latexsym,amsmath,amssymb,amsfonts,color}
\begin{document}

\title{Self-gravitating spinning mesonic tubes}

\author{Alex Giacomini}
\email{alexgiacomini@uach.cl}
\affiliation{Instituto de Ciencias F\'isicas y Matem\'aticas, Universidad Austral de Chile, Casilla 567, Valdivia, Chile.}
\author{Marcela Lagos}
\email{marcela.lagos@uss.cl}
\affiliation{Universidad San Sebastián, Avenida del Cóndor 720, Santiago, Chile.}
\author{Seung Hun Oh}
\email{sho@tukorea.ac.kr}
\affiliation{Department of Consilience, Tech University of Korea, Siheung 15073, Korea.}
\author{Aldo Vera}
\email{aldo.vera@umayor.cl}
\affiliation{N\'ucleo de Matem\'atica, F\'isica y Estad\'istica, Universidad Mayor, Avenida Manuel Montt 367, Santiago, Chile.}
\affiliation{Centro Multidisciplinario de F\'isica, Vicerrector\'ia de Investigaci\'on, Universidad Mayor, Camino La Pir\'amide 5750, Santiago, Chile.}

\begin{abstract}
In this paper, we construct new self-gravitating solutions for the four-dimensional $SU(2)$ Einstein non-linear sigma model and the Einstein-Skyrme model, both minimally coupled to the Maxwell field, which describes spinning pulsed mesonic tubes. These solutions are gyratonic in the sense that they are characterized by a propagation axis and possess angular momentum. The gyratonic solutions are described by a smooth matter distribution, which is free of curvature singularities. The maximum of the energy density is concentrated on a tube around the propagation axis. We first consider the $SU(2)$ non-linear sigma model minimally coupled to General Relativity and the Maxwell field, constructing asymptotically flat, spinning, pulsed gyratonic solutions. Subsequently, we include the Skyrme term in the analysis. In this case, we show that in order to maintain the resolvability of the matter field equations, the spacetime must be asymptotically de Sitter. It is also shown that, despite the fact that the total topological charge is zero, these solutions possess a non-zero topological charge density.
\end{abstract}

\maketitle

\newpage

\tableofcontents

\section{Introduction}
The first exact solution in General Relativity describing an ultrarelativistic spinless point particle is the well-known Aichelburg-Sexl metric \cite{AiSe}. The Aichelburg-Sexl metric is of great physical interest because it is conjectured that, in particle accelerators at extremely high energies, mini black holes may be formed in head-on collisions between particles. Using the Aichelburg-Sexl metric, the conditions for the formation of an apparent horizon in such collisions can be studied, as has been done in Refs. \cite{EaGi,YoNa1,YoNa2,YoRi}.
However, we know that all fundamental particles in nature, with the sole exception of the Higgs boson, are characterized by spin. Moreover, many composite subatomic particles relevant in high-energy physics, such as baryons, also possess spin. For this reason, in the calculations of mini-black-hole formation by particle collisions, the effects of spin-orbit and spin-spin interactions will lead to non-trivial physical effects that must be considered, as discussed in Ref. \cite{YoZe}.

To generalize the Aichelburg-Sexl metric to an exact solution describing a spinning particle, a natural approach is to apply a boost to the Kerr metric, analogous to how the Aichelburg-Sexl spacetime is obtained from the Schwarzschild metric (see Refs. \cite{FePe} and \cite{SaLo}). However, in the ultrarelativistic limit of this construction, the angular momentum vanishes. Therefore, to obtain an exact solution with finite angular momentum, one must instead consider a beam of spinning null fluid or circularly polarized electromagnetic radiation as a source. These solutions, known in the literature as gyratons, were first investigated in Ref. \cite{Bo} and independently rediscovered and analyzed in detail in Refs. \cite{FroFu}, \cite{FroIsZe} and \cite{FroZe1}. Later, the charged version of the gyratonic solution was described in Ref. \cite{FroZe2}.

The metric sourced by gyratons belongs to the Kundt class, a generalization of pp-waves, whose physical properties and exact relation to gyratons have been analyzed in detail in Ref. \cite{PoSte}. These exact solutions have since been extended to more generic backgrounds, including spacetimes with a cosmological constant, Robinson-Trautman geometries, and higher dimensions \cite{PoSva,KrPoZeKa1}. Furthermore, gyratonic configurations have been constructed in product spaces \cite{KrPoZeKa2} as well as in $2+1$ spacetime dimensions \cite{PoSvaMa}. Despite these developments, exact solutions in General Relativity describing spinning particle beams have only been studied in detail in these massless (or ultrarelativistic) cases.

The most consistent way to search for spinning matter beam sources is by deriving them from a fundamental action principle. For example, gyraton solutions have been explored in the context of supergravity in Refs. \cite{FroLi} and \cite{CaKleZo}. However, to date, in most cases, the matter fields supporting such solutions are described by an effective energy-momentum tensor not derived from an action principle. To provide the available spinning configurations, together with their intriguing properties (and with a robust foundation within the Standard Model), a mandatory task is to construct the energy-momentum tensor using ingredients from nuclear and particle physics.

In this context, the matter source is usually assumed to be localized around the propagation axis. The matter distribution can be concentrated on the axis and thus be singular, or it can be localized within a cylinder of finite radius. In the latter case, this implies that outside a cylindrical region, the vacuum Einstein field equations (possibly with a cosmological constant) apply: Such a solution describes a spinning matter tube. It is then possible to glue the inner solution with non-zero matter fields to the outer vacuum solution using suitable matching conditions. While using matter fields localized on a cylinder of finite radius avoids singularities, it is often difficult to provide an argument based on fundamental principles to estimate the cylinder's radius, which remains somewhat arbitrary. 

On the other hand, there are many physically interesting objects where the matter source has no sharp boundary but extends to infinity with sufficiently fast fall-off conditions, such as non-compact boson stars, where the matter field typically falls off exponentially (for a comprehensive review and its astrophysical relevance, see Ref.  \cite{Jetz}). In general, all solutions of field theories derived from a fundamental action principle, including non-self-gravitating cases, have fields extending to infinity, as they must be smooth to be considered physical solutions. It is, therefore, of great interest to generalize the concept of a spinning particle beam to cases where the matter field is non-zero outside a region of finite radius but falls off quickly enough and is free of singularities. This ensures that the solutions still possess finite angular momentum and energy per unit length. Furthermore, as a null fluid imposes strong constraints on the form of the stress-energy tensor, there is no \textit{a priori} reason to restrict the analysis only to massless or ultrarelativistic spinning particle beams. Dropping the hypothesis of a null fluid as a matter source can be seen as a natural generalization of the gyraton concept.

Typically, self-gravitating solutions describing matter beams or tubes (spinning or non-spinning) have an infinite extension in time and, consequently, infinite total energy and angular momentum, even if the values per unit length are finite. It would, therefore, be highly desirable to obtain an exact spinning matter tube solution where the property of compact support in time is compatible with the equations of motion. Additionally, the possibility that the spinning matter field source may possess a non-Abelian internal symmetry group has not yet been explored in the gyraton literature. 

Two phenomenologically successful field theories with non-Abelian internal symmetry are the non-linear sigma model and the Skyrme model (see, e.g., \cite{Nair, ManSu} and references therein). Both are effective low-energy theories of hadronic matter. In their simplest versions, the internal symmetry group is $SU(2)$, describing pions as elementary degrees of freedom and nucleons (neutrons and protons) as solitons.

Due to the highly non-linear nature of these models, finding exact solutions is extremely difficult even in flat spacetime. Indeed, it was only in 2013 that the first systematic technique to construct them for the Skyrme model was introduced \cite{Canf1}. Intuitively, one might expect that coupling these fields to General Relativity would make the task of finding analytic solutions excessively difficult. It thus came as a surprise that, in the same year, exact solutions for the Einstein-Skyrme theory were also found \cite{CanfMaed}. Subsequently, an exact hairy black hole solution with non-trivial thermodynamic behavior for the self-gravitating $SU(2)$ non-linear sigma model was discovered \cite{ACGO}. Furthermore, exact Bianchi IX cosmological solutions with topological charge are known for this model \cite{CDGP}, along with more generic self-gravitating configurations \cite{GO}.

Finding these analytic solutions was made possible by a generalized hedgehog ansatz. While the standard ansatz associates each point in physical space with a normal radial vector in isospin space, thus being spherically symmetric up to an internal isospin rotation, the generalization employed in the aforementioned papers replaces the radial profile and angular coordinates with more generic functions of the spacetime variables. Under this assumption, the full system of equations of motion (comprising the Einstein and matter field equations) simplifies drastically, as the latter are either identically satisfied or reduced to a single equation. This leads to the counter-intuitive situation where finding exact solutions for the non-linear sigma and Skyrme models is not only possible but actually easier in the self-gravitating case than in the non-gravitating one.
A crucial feature of this generalized hedgehog ansatz is that the matter field does not need to possess the same symmetries as spacetime since it enters the Einstein equations only via its energy-momentum tensor. In other words, the matter field can explicitly depend on Killing coordinates while its energy-momentum tensor does not. It is also worth noting that this ansatz can be adapted to various symmetries. For example, an interesting analytic solution found for the $SU(2)$ non-linear sigma model has the physical interpretation of a superconducting ``cosmic tube" (a thick cosmic string) \cite{CGLOV,Bonga:2022nwe}. These solutions have the maximum energy density of the matter field concentrated around the axis of the tube, and the density falls off rapidly enough to yield finite energy per unit length. However, the matter field remains non-zero outside a finite radius, and the solutions are free of curvature invariant singularities.

In the present paper, we generalize the solutions presented in \cite{CGLOV,Bonga:2022nwe} to incorporate gyratonic effects, which are naturally expected in hadronic processes involving pions and baryons. In high-energy hadronic collisions, particles often propagate at ultrarelativistic speeds, where both their intrinsic spin and the orbital angular momentum of the system play a crucial role. Since the Skyrme and non-linear sigma models describe hadrons as solitons with a well-defined spatial extension, their propagation in the ultrarelativistic limit cannot be fully captured by the spinless Aichelburg-Sexl-type metric. By including non-Abelian degrees of freedom, we provide a more realistic description of the internal symmetries of nuclear matter, where the coupling between spin and the gravitational field offers a richer framework for studying gravitational back-reaction and the formation of apparent horizons in hadronic jets.

We demonstrate that these novel gyratonic solutions, supported by charged meson matter fields, can be constructed in a straightforward manner from fundamental action principles. First, we employ the $SU(2)$ non-linear sigma model to derive gyratonic configurations in asymptotically flat spacetimes. Then, we generalize this construction to the $SU(2)$-Skyrme model, showing that, in this scenario, the spacetime metric must be asymptotically de Sitter to maintain the resolvability of the field equations. To achieve these results, we use recent developments in the construction of self-gravitating solitons, black holes, and extended objects within the Einstein-Skyrme framework; see \cite{Ayon-Beato:2015eca,Ayon-Beato:2019tvu,ACGO,Astorino:2018dtr,Henriquez-Baez:2022ubu,Concha:2023qcs,Henriquez-Baez:2024rjb,Vera:2025qqz,Canfora:2026wfi,Canfora:2026dym,Canfora:2026col} and references therein.\footnote{It is also worth emphasizing that using the results in \cite{BPS1,BPS2,BPS3,BPS4,BPS5,BPS6} might make it possible to extend the present results to include further matter fields (such as complex scalar fields describing superfluid condensates). In fact, this intriguing issue is under investigation.}

The paper is organized as follows. Section II introduces the gauged Einstein-Skyrme model, proposes a suitable ansatz for the metric and matter fields, and derives the field equations. Section III presents an exact, asymptotically flat, pulsed gyraton solution for the gauged $SU(2)$ Einstein-NLSM. We demonstrate that this solution is free of singularities without requiring fine-tuning. In Section IV, we generalize this configuration by including the Skyrme term, showing that it forces the spacetime to be asymptotically de Sitter. In Section V, we will compute the angular momentum for both the asymptotically flat and asymptotically de Sitter solution. In Section VI, we will analyze  the energy density and topological charge density. Section VII will be dedicated to the conclusions.

\section{Preliminaries}

\subsection{The gauged $SU(2)$ Einstein-Skyrme model}

The four-dimensional $SU(2)$ Einstein-Skyrme model minimally coupled to an abelian $U(1)$ Maxwell field is described by the action
\begin{equation}
I[g, U, A]=\int_{\mathcal{M}} d^{4}x\sqrt{-g}\left( \frac{R-2\Lambda }{2\kappa } + \frac{K}{4}\mathrm{Tr}[L^{\mu }L_{\mu }]+\frac{K\lambda }{32}\mathrm{Tr}\left( G_{\mu \nu }G^{\mu \nu }\right)  -\frac{1}{4}F_{\mu \nu }F^{\mu \nu } \right) \ ,  \label{I}
\end{equation}%
where $g$ denotes the determinant of the spacetime metric $g_{\mu\nu}$, $R$ is the Ricci scalar, and $\Lambda$ represents the cosmological constant. The parameters $K$ and $\lambda$ are positive coupling constants, and the electromagnetic field strength tensor is defined as $F_{\mu\nu} = \partial_\mu A_\nu - \partial_\nu A_\mu$. The matter field $U(x)$ takes values in the $SU(2)$ Lie group, whose generators are defined via the Pauli matrices $\sigma_a$ as $t_{a}= i \sigma_a$. The gauged Maurer-Cartan left-invariant current $L_{\mu}$ and the associated Skyrme curvature tensor $G_{\mu\nu}$ are given by
\begin{equation}
L_{\mu }=U^{-1}D_{\mu }U=L_{\mu }^{a}t_{a}\ , \qquad G_{\mu \nu }=[L_{\mu},L_{\nu }]\ .
\end{equation}%
Here, $D_{\mu }$ denotes the covariant derivative associated with the $U(1)$ gauge potential $A_{\mu }$, which acts explicitly on the $SU(2)$ field as
\begin{equation}
D_{\mu }U=\nabla _{\mu }U+A_{\mu }U\hat{O}\ ,\qquad \hat{O}=U^{-1}\left[ t_{3},U\right] \ ,  \label{covdev}
\end{equation}%
where $\nabla_\mu$ denotes the standard Levi-Civita covariant derivative, and reduces to the standard partial derivative when acting on the group element.

Varying the action in Eq. \eqref{I} with respect to the gravitational, Skyrme, and gauge fields yields the complete system of coupled field equations:
\begin{gather}
R_{\mu \nu }-\frac{1}{2}g_{\mu \nu } R+\Lambda g_{\mu \nu }=\kappa T_{\mu \nu}\ , \label{Eq2}\\
D_{\mu }L^{\mu }+\frac{\lambda }{4} D_{\mu }[L_{\nu },G^{\mu \nu}]=0\ , \label{Eq1}\\
\nabla _{\mu }F^{\mu \nu } =J^{\nu } \ , \label{Eq3} 
\end{gather} 
where the conserved electromagnetic current $J^{\mu}$ is given by
\begin{equation}
J_{\mu }=- \frac{K}{2}\text{Tr}\Bigr[\hat{O}\Bigr(L_{\mu }+\frac{\lambda }{4}[L^{\nu },G_{\mu \nu }]\Bigr)\Bigr] \ .
\end{equation}
The total energy-momentum tensor $T_{\mu \nu }$ in Eq. \eqref{Eq2} decomposes into the sum of the Skyrme and Maxwell contributions, respectively:
\begin{equation}
T_{\mu \nu }= T_{\mu\nu}^{\text{Sk}} + T_{\mu\nu}^{\text{Max}} \ ,  \label{Tmunu}
\end{equation}
where the individual energy-momentum tensors are defined as
\begin{align}
T_{\mu\nu}^{\text{Sk}} &= -\frac{K}{2}\mathrm{Tr}\left[ L_{\mu }L_{\nu }-\frac{1}{2}g_{\mu \nu }L^{\alpha }L_{\alpha } + \frac{\lambda }{4}\left( g^{\alpha \beta }G_{\mu \alpha }G_{\nu \beta }-\frac{g_{\mu \nu }}{4}G_{\sigma \rho }G^{\sigma \rho }\right) \right] \ , \\
T_{\mu \nu }^{\text{Max}} &=g^{\alpha \beta }F _{\mu \alpha}F_{\nu \beta }-\frac{1}{4}g_{\mu \nu }F^{\alpha \beta}F_{\alpha \beta } \ .
\end{align}

\subsection{Matter field and geometric setup}

In order to construct a self-gravitating spinning tube solution in the $SU(2)$ Einstein-Skyrme model, we model the four-dimensional spacetime using a half-null (characteristic) coordinate system $x^\mu = \{u, v, r, \phi \}$, where the spatial coordinates span the ranges $0 < r < \infty$ and $0 \le \phi < 2\pi$. A compatible metric ansatz within the Kundt class is chosen as follows:
\begin{equation}  \label{metric}
 ds^2 = -f du^2 +e^{R_1}(dr^2+ r^2d\phi^2)  +2 dudv + 2 R_2 dud\phi  \ ,
\end{equation}
where the profile functions satisfy $f=f(u,r)$, $R_1=R_1(r)$, and $R_2=R_2(u,r)$. Throughout this work, $f$ is assumed to be positive-definite so that the coordinate $u$ represents a timelike evolution parameter ($g_{uu} = -f < 0$), whereas $v$ designates a strictly null coordinate running along the generators ($g_{vv} = 0$). 

The surfaces of constant $u$, denoted by $\Sigma_u$, define a one-parameter family of three-dimensional null hypersurfaces whose normal one-form is given by $\ell_\mu = \partial_\mu u = \delta^u_\mu$. The cross-metric normalization constraint $g_{uv} = 1$ dictates that the corresponding contravariant vector field, $\ell^\mu = g^{\mu\nu}\ell_\nu = (\partial/\partial v)^\mu$, is simultaneously normal and tangent to $\Sigma_u$, thereby serving as the generator of the lightlike trajectories. Crucially, all metric components are invariant under translations along these null generators ($\partial_v g_{\mu\nu} = 0$), implying that $\chi^\mu = (\partial/\partial v)^\mu$ constitutes a strictly null Killing vector field everywhere. The remaining coordinates, $(r, \phi)$, span the two-dimensional spacelike cross-sections possessing a polar disk topology characterized by a positive-definite, orthogonal metric block ($g_{rr} = e^{R_1} > 0$, $g_{\phi\phi} = e^{R_1}r^2 > 0$,  $g_{r\phi} = 0$). Axial symmetry is generated by the rotational Killing vector field $\xi^\mu = (\partial/\partial \phi)^\mu$, which couples to the time evolution via the non-vanishing frame-dragging component $g_{u\phi} = R_2 \neq 0$. Under this specific geometric gauge, the inverse metric components strictly satisfy $g^{uu} = 0$, $g^{ur} = 0$, and $g^{u\phi} = 0$. This ensures that the full four-dimensional volume determinant factorizes cleanly into the transverse spatial measure, $\sqrt{-g} = \sqrt{g_{rr}g_{\phi\phi}} = r e^{R_1}$, providing an algebraic isolation that simplifies integration over the characteristic slice.

It is important to stress that the metric ansatz is not exactly that of a gyraton because of the non-trivial $R_1$ function in front of the transverse metric. Indeed, the existence of this non-constant function implies that the energy-momentum tensor sourcing this metric acquires additional components that do not correspond to a null fluid. However, the metric is still characterized by the non-diagonal gyraton-like term $g_{u\phi}=R_2$. This off-diagonal term cannot be gauged away by a coordinate transformation if the two-form $\mathbf{\Omega}=(R_2)_{,r}\mathbf{d}r\wedge\mathbf{d}\phi\neq 0$. Indeed, it was shown in Ref. \cite{PoSte} that a non-zero $\mathbf{\Omega}$ is responsible for frame-dragging. It is also worth stressing that for standard gyratons, if the matter source is concentrated on a tube of finite radius $R_0$, then for $r>R_0$, the $T_{u\phi}$ component of the energy-momentum tensor is zero, implying that the two-form $\mathbf{\Omega}$ is also zero. Therefore, at least locally, the $g_{u\phi}$ component of the metric can be gauged away. However, because the vacuum domain is not simply connected, it cannot be globally gauged away, leading to a non-trivial angular momentum. In our case, as the matter field has no sharp boundary, the $g_{u\phi}$ component of the metric cannot be gauged away, not even locally.

To support this stationary, axisymmetrically dragged background, we consider a matter configuration representing inhomogeneous condensates. In the Euler angles parametrization, the $SU(2)$ group element ansatz for the Skyrme field maps to crystals of hadronic layers \cite{SU(N)1,SU(N)2,SU(N)3} and is written as:
\begin{equation} \label{U1}
 U= e^{F_1(x^\mu) \cdot t_3} e^{F_2(x^\mu) \cdot t_2} e^{F_3(x^\mu) \cdot t_3}  \ , 
\end{equation}
where the three internal degrees of freedom are distributed across the coordinates according to:
\begin{equation} \label{U2}
 F_1(x^\mu) = \frac{p}{2} \phi \ , \qquad F_2(x^\mu) =  H(r) \ , \qquad F_3(x^\mu) = G(u) \ ,
\end{equation}
for a winding constant $p$. Concurrently, the ansatz for the Maxwell potential $A_\mu$ is chosen to match the target symmetries:
\begin{equation} \label{Amu}
 A_\mu dx^{\mu} = B(u,r) du- \frac{p}{4} d\phi \ .
\end{equation}
As the spacetime metric is independent of the null coordinate $v$, both matter fields are assumed to be independent of $v$, ensuring full compatibility with the underlying field equations. 

While the global energy naturally reflects the infinite extent of the tube geometry, a well-defined and physically meaningful measure is obtained by computing the energy traversing null hypersurfaces per unit affine length, as formulated in Refs. \cite{FroFu}, \cite{FroIsZe}:
\begin{equation}  \label{energy}
\mathcal{E} = - \int_{\Sigma} \sqrt{g_{2}} T^{\mu}_{\phantom{\mu}u} d\Sigma_{\mu}
= -2\pi \int_{r} r e^{R_1} T^{u}_{\phantom{u}u} \, dr \ .
\end{equation}

\subsection{Field equations}

The first step to solving the system of equations is to realize that 
with the above ansatz, the function $G(u)$ can be removed from the complete system of equations, considering that 
\begin{gather}
    B(u,r)=\frac{1}{2}\dot G(u) (1-P(r)) \ , \\ 
    R_2(u,r)= \dot G(u) V(r) \ , \\ 
    f(u,r)= (\dot G(u))^2 M(r) \ ,
\end{gather}
where we have used the notation: $\dot{} \equiv \partial_u$ and ${}^\prime{} \equiv\partial_r$. This change leads to a system of seven coupled ordinary differential equations, divided as follows:
\begin{enumerate}
\item The Skyrme equations reduce to a single ODE:
\begin{align}
\lambda (2-p\sin(H)+r R_1') = 0 \ .
\label{skeq}
\end{align}
\item The Maxwell equations reduce to a single ODE:
\begin{align}
P'' + \frac{1}{r} P' - 4KP \sin^{2}H \, \bigg[
 e^{R_{1}} + \lambda \bigg(H^{\prime 2} + \frac{p^{2}}{4r^{2}}
\cos^{2}H \bigg)\bigg]=0 \ .
\label{maxeq}
\end{align}
\item Einstein's equations are
\begin{align}
(H')^2 - \frac{p^{2}}{4r^{2}} \cos^{2} (H) = 0 \ ,\label{constraint1} \\
 K\kappa \lambda p^{2} \cos^{2} (H)  \, (H')^2
- 8  \Lambda r^{2} e^{2R_{1}} = 0 \ , \label{constraint2} \\
R_{1}^{\prime \prime} + \frac{1}{r} R_{1}^{\prime} + \frac{K \kappa p^{2}}{2r^{2}} \cos^{2} (H) 
+ \frac{K \kappa \lambda p^{4}}{16 r^{4}} \cos^{4} (H) e^{-R_{1}}+ 2 \Lambda e^{R_{1}} = 0 \ ,  \label{eqR1}\\
V'' - \bigg( R_{1}^{\prime} + \frac{1}{r} \bigg) 
V' + K \kappa \, p \, e^{R_{1}} \cos^{2} (H)
+ \frac{K\kappa \lambda p^3}{4r^2}\cos^{4} (H) = 0 \ , \label{eqR2}\\
M'' + \frac{1}{r} M' + \frac{e^{-R_{1}}}{r^{2}} 
\big(  V' \big)^{2} - \frac{\kappa}{2} \big( P' \big)^{2} - 2K\kappa e^{R_{1}} \big( P^{2} \sin^{2}(H) + \cos^{2}(H)
\big) \notag \\
- \frac{K\kappa \lambda p^2}{4 r^2} \bigg[ P^{2} \sin^{2}(2H)+ 2\cos^{4}(H) \bigg] 
= 0  \ . \label{eqf}
\end{align}
\end{enumerate}
Here we can note some interesting facts. First, the Skyrme equation is, in principle, a second-order differential equation for the Skyrme profile, which is given by
\[
\biggl( e^{R_1} + \frac{p^2\lambda}{4r^2}\cos^2(H) \biggl)H'' +\biggl( \frac{e^{R_1}}{r} - \frac{p^2\lambda}{4r^3} \cos^2(H)(1+rR'_1) \biggl)H' -\frac{p^2\lambda}{8r^2} \sin(2H)(H')^2 + \frac{p^2}{8r^2}e^{R_1}\sin(2H)      = 0 \ . 
\]
However, this equation can easily be put in the form of Eq. \eqref{skeq} using the Einstein equations mentioned above.
Second, Eq. \eqref{constraint1} can be directly solved, obtaining 
\begin{equation} \label{Hsol}
    H = \arcsin\biggl( \frac{c_1 r^{p} - 1}{c_1 r^{p} + 1}  \biggl) \ ,
\end{equation}
where $c_{1}$ is a positive constant. Note that the solution for the case $p<0$ can be written in the same form as Eq. (26) with the overall sign flipped. Since the governing differential equations depend only on $(H')^2$ and $\cos^2(H)$, the sign flip leaves the equation invariant. Thus, without loss of generality, we restrict our detailed analysis to $p > 0$. As we will discuss later, for larger values of $c_{1}$ and $p$, the energy of the Skyrme field will become increasingly localized in a narrower region around the axis of symmetry. 

On the other hand, Eq. \eqref{constraint2} leads to two different branches; the first one is to consider $\Lambda=\lambda=0$, and the second is to solve the $R_1$ function in terms of $H$ for non-vanishing values of the coupling constants.
In the next sections, we will show that the above situations reduce the $SU(2)$ Einstein-Skyrme equations, allowing us to construct two different families of solutions.

\section{Asymptotically flat spinning mesonic tubes}

First, let us consider the case of an asymptotically flat space-time. It is a direct computation to verify that the Einstein equation in Eq. \eqref{constraint2} demands that the Skyrme term vanish. Then, we are in the sector
\begin{equation}
 \Lambda=0 \ , \qquad \lambda=0 \ .
\end{equation}

\subsection{Constructing regular solutions}

In the present case, Eq. \eqref{eqR1} is reduced to
\begin{align}
R_1'' + \frac{1}{r} R_1' + \frac{2K \kappa p^2 c_{1} r^{p}}
{r^2 \big( 1+ c_{1}r^{p} \big)^{2} }  = 0 \ , 
\end{align}
and the solution to this equation is found to be
\begin{align}
 R_1 = (2K \kappa p +c_2) \log(r) -2 K \kappa \log(1+c_1 r^{p}) + c_3\ , \label{R1sol1}
\end{align}
for some constants $c_{2}$ and $c_{3}$. Note that the constant $c_3$ can be eliminated by a suitable rescaling of the coordinate $r$; in what follows, we will consider $c_3=0$. To examine the regularity on the symmetry axis, 
let us introduce a new coordinate, $X$, defined by
\begin{equation*}
X = \frac{2}{m} r^{m/2} \ , \qquad \textrm{where} \qquad
m = 2 (K\kappa p+1) + c_{2} \ .
\end{equation*}
Then, the induced metric $ds_{2}^{2}$ on the 2-surface with constant $u$ and $v$ 
approximates to
\begin{equation*}
ds_{2}^{2} \approx dX^{2} + \frac{m^{2}}{4} X^{2} d\phi^{2} \ ,
\end{equation*}
near the symmetry axis. To remove the conical singularity along the symmetry axis, we should impose $m=2$, which means $c_{2} = - 2K \kappa p$. Then, the $R_1$ function and the metric component $g_{rr}$ become
\begin{equation} \label{R1sol}
R_1 = -2 K \kappa \log(1+c_1 r^{p}) \ , \qquad g_{rr} = e^{R_{1}} = \frac{1}{\big( 1+c_{1}r^{p}\big)^{2K\kappa}} \ .
\end{equation}
To examine the angle defect at spatial infinity, let us introduce a radial coordinate $X_{\infty}$ given by
\begin{align}
     X_{\infty} 
     = \frac{c_{1}^{-K\kappa}}{1-K\kappa p} r^{1-K\kappa p} \ .
\end{align}
Then, the two-dimensional induced metric on the transverse section at spatial infinity can be written in terms of $X_{\infty}$ as
\begin{align}
    ds_{2}^{2} \big\vert_{r=\infty} 
    = dX_{\infty}^{2} + \big( 1-K\kappa p \big)^{2} X_{\infty}^{2} \,
    d\phi^{2} \ .
\end{align}
This demonstrates that the spacetime exhibits an angle defect at spatial infinity proportional to $p$, thereby behaving asymptotically as a cosmic string. 
With the explicit form of $R_{1}$, we can rewrite Eqs. \eqref{maxeq} and \eqref{eqR2} as
\begin{align}
& P'' + \frac{1}{r} P' - \frac{4K \big(c_{1}r^{p}-1 \big)^{2}}
{\big(c_{1}r^{p}+1 \big)^{2+2K\kappa}} P = 0 \ , 
\label{EqP1} \\
& V^{\prime \prime} - \bigg[ \frac{1- 2K \kappa p}{r} 
+ \frac{ 2K \kappa p}{r \big(1+c_{1}r^{p} \big)} \bigg] V^{\prime}
+ \frac{  4 K \kappa p c_1 r^{p}}{\big(1+c_{1}r^{p} \big)^{2(K\kappa +1)}} = 0 \ . \label{eqV} 
\end{align}
The general solution of Eq. \eqref{eqV} is given by a linear combination of hypergeometric functions:
\begin{align} \label{Vsol1}
V= 2 K \kappa \, r^{2} {}_{2}F_{1} (1+2K\kappa, 2/p,
1+2/p; -c_{1}r^{p})  
+ V_{1} r^{2} {}_{2}F_{1} (2K\kappa, 2/p,1+2/p; -c_{1}r^{p}) + V_{2} \ ,
\end{align}
where $V_{1}$ and $V_{2}$ are integration constants fixed by regularity conditions at $r=0$ and $r=\infty$. Near the symmetry axis $r=0$, $V$ is expanded as
\begin{align}
    V = V_{2} + \big( 2K \kappa + V_{1} \big) r^{2} + O(r^{p+2}) \ .
\end{align}
To prevent conical singularities along the symmetry axis, we must set $V_{2} = 0$. While setting $V_{2}=0$ ensures $V(0)=0$, $V^{\prime}(0)=0$ holds automatically regardless of the values of the integration constants. The remaining integration constant $V_{1}$ is fixed by requiring the angular velocity of Zero Angular Momentum Observers (ZAMOs), $\omega = -g_{u\phi}/g_{\phi\phi}$, to remain finite everywhere. For $K\kappa p \le 1$, the second term in Eq. \eqref{Vsol1} causes $\omega$ to diverge unless $V_{1}=0$. On the other hand, for $K\kappa p > 1$, finiteness of $\omega$ fixes $V_{1}$ to
\begin{align}
    V_{1} = - \frac{4 (K\kappa p -1)}{p^{2} c_{1}^{4/p}} \Gamma(2/p) \ .
    \label{V1largep}
\end{align}
In contrast to the metric function $V(r)$, the Maxwell field $P(r)$ cannot be accommodated for arbitrary values of $K\kappa p$. In the asymptotic limit $r \to \infty$, the differential equation for $P(r)$ reduces to
\begin{align}
\frac{d^{2}P}{dr^{2}} + \frac{1}{r} \frac{d P}{dr} 
- \frac{4K c_{1}^{-2K\kappa}}{r^{2K\kappa p}} P = 0 \ .
\label{asymp_P}
\end{align}
By introducing the coordinate transformation
\begin{align}
\rho(r) \equiv \frac{2\sqrt{K}}{c_{1}^{K\kappa} |K\kappa p - 1|} r^{1-K\kappa p} \ ,
\end{align}
Eq. \eqref{asymp_P} transforms into the modified Bessel differential equation of order zero,
\begin{align}
\rho^{2} \frac{d^{2}P}{d\rho^{2}} + \rho \frac{dP}{d\rho} - \rho^{2} P = 0 \ ,
\end{align}
whose general solution is a linear combination of modified Bessel functions:
\begin{align}
P(\rho) = p_{\infty}^{(1)} I_{0} (\rho) + p_{\infty}^{(2)} K_{0} (\rho) \ .
\end{align}
Note the mapping of the boundaries: when $K\kappa p < 1$, the physical limits $r=0$ and $r \to \infty$ correspond to $\rho=0$ and $\rho \to \infty$, respectively; conversely, when $K\kappa p > 1$, they correspond to $\rho \to \infty$ and $\rho=0$, respectively. Because $I_0(\rho)$ diverges as $\rho \to \infty$ and $K_0(\rho)$ diverges as $\rho \to 0$, enforcing regularity at both spatial boundaries in either parameter regime requires $p_{\infty}^{(1)} = p_{\infty}^{(2)} = 0$. Therefore, a non-trivial Maxwell field is allowed only when $K\kappa p = 1$. 

Now, let us concentrate on the cases with $K\kappa p=1$. In order to compactify the radial coordinate, we introduce a new variable, $z$, defined by
\begin{equation}
    z(r) = \frac{c_{1} r^{p}}{1+ c_{1} r^{p}} \ .
\end{equation}
This monotonic transformation maps the physical domain $r\in [0, \infty)$ into the bounded interval $z\in [0,1)$. In terms of $z$, Eq. \eqref{EqP1} can be recast as
\begin{align}
    \frac{d^{2}P}{dz^{2}} + \frac{1-2z}{z(1-z)} \frac{dP}{dz} 
    - \frac{4K}{p^{2}c_{1}^{2/p}} \frac{z^{2/p}(2z-1)^{2}}{z^{2}(1-z)^{2}}P
    =0 \ .
\end{align}
For arbitrary $p > 2$, the fractional power $z^{2/p}$ in the final term introduces a branch point at $z=0$, which is not desirable. While the formal generalized series solution for this arbitrary case is detailed in the Appendix \ref{AppB}, in this section we focus on the specific case of $K\kappa=1/2$ and $p=2$, which permits an exact analytical solution in closed form. To factor out the asymptotic behavior, we introduce the transformation
\begin{align}
P(z) = (1-z)^{\sigma} Y(z) \ ,
\end{align}
where the characteristic exponent is $\sigma = \pm \sqrt{\frac{K}{c_{1}}}$.
Substituting this transformation, we find that the function $Y(z)$ satisfies the equation
\begin{align}
\bigg[\frac{d^{2}}{dz^{2}} +\frac{1-2(\sigma +1) z}{z(1-z)}\, \frac{d}{dz}
+ \frac{\sigma (4 \sigma z - \sigma -1)}{z(1-z)}  \bigg] Y(z)=0 \ .
\end{align}
This can be mapped directly to the canonical form of the confluent Heun equation \cite{Heun}
\begin{align}
Z^{\prime\prime} + \bigg(\frac{\gamma}{z} + \frac{\delta}{z-1} + \epsilon \bigg)Z^{\prime}
+ \frac{\alpha z - q}{z(z-1)} Z = 0 \ ,
\end{align}
whose regular solution is conventionally denoted by
\begin{align}
Z = \textrm{HC}(q, \alpha, \gamma, \delta, \epsilon ; z) \ .
\end{align}
The confluent Heun equation admits two linearly independent solutions, yielding the general form
\begin{align}
Z = z_{1} \textrm{HC}(q, \alpha, \gamma, \delta, \epsilon ;z) + z_{2} (z-1)^{1-\gamma} \textrm{HC}(q - (1-\gamma)\delta + (1-\gamma)\epsilon, \,\, \alpha + (1-\gamma)\epsilon, \,\, 2-\gamma, \,\, \delta, \,\, \epsilon, \,\, ;z) \ ,
\end{align}
for some constants $z_{1}$ and $z_{2}$. Because $\gamma=1$ in our parameterization, the second term exhibits a logarithmic divergence. We discard this nonphysical branch to maintain regularity by setting $z_{2}=0$. Thus, the exact global solution to our equation is given by 
\begin{align}
P(z) = p_{0} (1-z)^{\sigma} \, \textrm{HC}\big(-\sigma (\sigma+1) ,-4\sigma^{2}, 1, 
2\sigma+1, 0; z \big) \ ,
\end{align}
where $p_{0} = P(0)$ is the core amplitude. To satisfy the physical boundary condition at $r=\infty$, we select the positive root for $\sigma$; that is $\sigma = \sqrt{\frac{K}{c_{1}}}$. This choice guarantees that $P(1)=0$, ensuring that the Maxwell field properly vanishes at spatial infinity.

Now, we are left with the equation for $M$:
\begin{align}
 M'' + \frac{1}{r} M' - \frac{\kappa}{2}(P')^2    - 2K \kappa e^{R_1}(\cos^2(H) +2P^2 \sin^2(H) ) + \frac{e^{-R_1}}{r^2} (V')^2  = 0 \ . \label{EqM1}
\end{align}
It should be emphasized that the curvature invariants of our spacetime are simply polynomials of the Ricci scalar that is given by 
\begin{align}
    R = \frac{2 K \kappa p^2 c_1  r^{p-2}}{(1+c_1 r^{p})^{2(1-K \kappa)}} \ .
\end{align} 
To ensure the regularity on the symmetry axis, we impose $p \geq 2$ throughout this paper. At spatial infinity, the curvature scalar is approximately given by 
\begin{align}
R \sim \frac{2 K \kappa p^2}{c_1^{1-2 K \kappa}} \cdot \frac{1}{r^{2+(1-2K \kappa) p}} \ . 
\end{align}
Since the curvature invariants do not depend on the functions $P$ and $M$, the regularity of the solutions is guaranteed by the conditions on $p$ imposed above. Thus, once the solution of Eq. \eqref{EqP1} and \eqref{EqM1} are obtained, the system is completely solved  describing self-gravitating mesonic tubes with nonzero angular momentum (as we will see below). At this point, the equation for $P$ is approximately solved with high precision in the range up to $r\approx \exp(1/K\kappa p)$, and the corresponding $M$ function is obtained numerically. Also, when the Maxwell field is turned off, the equation for $M$ is solved analytically, and the condition that guarantees the positivity of $M$ is provided. 

In the next subsection, we will show that there is a particular sector in which the system can be solved only using analytical methods.

\subsection{An exact solution without the electromagnetic coupling}

In the absence of the electromagnetic coupling (also, $\Lambda=\lambda=0$), an exact analytical solution can be found by setting the function $P$ to zero. The remaining equation for $M(r)$ then reduces to
\begin{equation*}
   \frac{1}{r}\big( rM^{\prime} \big)^{\prime} = \frac{8K \kappa c_{1}r^{p}}
   {\big(c_{1}r^{p} +1 \big)^{2+2K\kappa}}
   -\frac{4}{\big(c_{1}r^{p} +1 \big)^{2K\kappa}} \bigg(\frac{2K\kappa}{c_{1}r^{p}+1} + V_{1}\bigg)^{2} \ ,
\end{equation*}
whose solution has the form: 
\begin{align} 
M =& -\frac{ 4 r^{2}}{p^{2} \big( 1+c_{1}r^{p}\big)^{2K\kappa}}
+m_{1} \, r^{2}  \, _{2}F_{1}\big(2K\kappa, 2/p;2/p+1; -c_{1}r^{p} \big) 
\notag \\
& - \frac{p^{2}m_{1}^{2}}{16} \, r^{2}
\, _{3}F_{2}\big(2K\kappa,2/p, 2/p;2/p+1, 2/p+1; -c_{1}r^{p} \big) + M_{0} \ ,  \label{Msol1}
\end{align}
where $M_{0}$ is an integration constant, and the coefficient $m_{1}$ is given by
\begin{align}
m_{1} 
=  \frac{4}{p} \left( \frac{2}{p} - 2K\kappa - V_1 \right) \ .
\end{align}
Note that one of the integration constants is fixed by imposing the standard regularity conditions at the axis, $M^{\prime}(0)=0$. Expanding the exact solution at spatial infinity, $M$ takes the asymptotic form:
\begin{align}
  M = M_{\infty} + M_{\textrm{ln}} \ln (c_{1} r^{p}) 
  + M_{\textrm{pow}} r^{2-2K\kappa p}+ O(r^{-\nu}) \ ,
\end{align}
where $\nu>0$. The first term represents the effective background constant, given by
\begin{align}
    M_\infty =  M_{0} + M_{\ln} \left[ \psi(2K\kappa - 2/p) - \psi(2/p) + \frac{2}{2K\kappa - 2/p + V_1} \right]  \ ,
\end{align}
where $\psi(x)$ is the digamma function. The second and third terms represent potential asymptotic divergences, with coefficients given by 
\begin{align}
    M_{\textrm{ln}} = - \frac{m_{1}^{2}\, \Gamma\left(2K\kappa - 2/p\right)\Gamma\left(2/p\right)}{4 c_1^{2/p}\Gamma(2K\kappa)}  \ , \qquad
    M_{\textrm{pow}} = -\frac{V_{1}^{2} c_{1}^{-2K\kappa}}
    {\big( 1- K\kappa p \big)^{2}} \ .
\end{align}
As discussed previously, the asymptotic behavior of $M$ is highly sensitive to the physical parameters. Let us examine each parametric regime in turn.

\vspace{.6cm}

\textbf{Case 1: $K\kappa p <1$}

\vspace{.6cm}

As discussed in the previous section, to ensure a finite angular velocity for zero-angular-momentum observers (ZAMOs) when $K\kappa p <1$, we must set $V_{1}=0$. Imposing this condition yields
\begin{equation}
    M_{\textrm{ln}} > 0 \ , \qquad M_{\textrm{pow}} = 0 \ , \qquad
    \nu = (1+ 2K \kappa) p- 2 > 0 \ ,
\end{equation}
where we have used the relations $-1<2K\kappa - 2/p<0$ and $p\geq 2$. The appearance of the logarithmic divergence in this result can be understood in terms of the Newtonian limit. For an observer at spatial infinity, the energy distribution along an infinite tube is perceived as an infinitely long wire, whose corresponding Newtonian potential is simply $\ln(r)$ multiplied by a suitable constant.

\vspace{.6cm}

\textbf{Case 2: $K\kappa p >1$}

\vspace{.6cm}

In this regime, $V_{1}$ is fixed by Eq. \eqref{V1largep} to avoid an infinite frame-dragging effect at spatial infinity. In the large-$r$ limit, $M$ behaves asymptotically as
\begin{equation}
M = M_{\infty} + M_{\textrm{ln}} \ln(r) + O(r^{\nu}) \ ,
\end{equation}
where $\nu=\textrm{max}(2-2K\kappa p, -p)$ . In this case, since $M_{\textrm{ln}}<0$, $M$ goes to $-\infty$ at spatial infinity, meaning it must change its sign at some finite distance. Notice that this does not imply the existence of an event horizon. Since the spacetime is spinning, this root merely defines the boundary of an ergoregion where no stationary observer can exist.

\vspace{.6cm}

\textbf{Case 3: $K\kappa p=1$}

\vspace{.6cm}

In this case, setting $V_{1}=0$ to ensure finite angular velocities for co-rotating observers at spatial infinity implies $m_{1}=0$. This straightforwardly yields the analytic solution
\begin{equation}
M= 1 - \frac{4r^{2}}{p^{2} \big(c_{1}r^{p} +1 \big)^{2/p}} \ ,
\end{equation}
where we have chosen $M_{0}=1$ to guaranty asymptotic flatness. Since this function is monotonically decreasing, $M(r)$ is positive definite if the following inequality holds:
\begin{equation}
   c_{1} > \left( \frac{2}{p} \right)^{p}\ . 
\end{equation}
Due to the non-trivial $R_1$ function and consequently the presence of non-vanishing fluid components in the energy-momentum tensor, this solution can be seen as a generalized gyraton. Indeed, we show in Section V that the solution is characterized by both a propagation axis and an angular momentum.

\section{Asymptotically de Sitter spinning mesonic tubes}

In this section, we will consider the case $\lambda\neq 0$, $\Lambda\neq 0 $. As we have pointed out before, Eq. \eqref{constraint2} can also be satisfied by fixing the function $R_1$ in terms of $H$ for non-vanishing coupling constants, with $H$ still given by Eq. \eqref{Hsol}. 

\subsection{Constructing regular solutions}

Let us consider again the metric in Eq. \eqref{metric}, the matter field in Eq. \eqref{U2} and the Maxwell potential in Eq. \eqref{Amu}. Eq. \eqref{constraint2} can be directly solved by $R_1$, obtaining 
\begin{align}
R_{1} = \ln \bigg( \sqrt{\frac{K\kappa\lambda}{32\Lambda}} \frac{p^{2}}{r^{2}}
\cos^{2} (H) \bigg) \ .
\label{R1itoH}
\end{align}
The function $H$ is given by Eq. \eqref{Hsol}, with a range 
\begin{align}
H(r) \, : \, - \frac{\pi}{2} \longrightarrow   \frac{\pi}{2} \qquad \textrm{as}
\qquad r \, : \, 0 \longrightarrow \infty \ ,
\end{align}
so that, $\cos (H) \geq 0 $. One can notice that Eq. \eqref{eqR1} is compatible with the above expression if the following constraint between the coupling constants is satisfied 
\begin{equation} \label{constraintcouplings}
K \kappa + \sqrt{2\Lambda \lambda} \sqrt{K\kappa} - 1 = 0 \quad \Rightarrow  \quad \lambda = \frac{(1-K\kappa)^2}{2K \kappa \Lambda} \ ,
\end{equation}
for positive values of the cosmological constant.
This constraint determines the value of the Skyrme coupling, which is of the same order as the value that allows the existence of the self-gravitating skyrmions; see Refs. \cite{Ayon-Beato:2015eca} and  \cite{Ayon-Beato:2019tvu}.

Now, let us analyze the regularity of the metric. The metric component $g_{rr}$ is given by
\begin{equation*}
g_{rr} = e^{R_{1}}=
\frac{c_{1}p^{2}(1-K\kappa)}{2\Lambda} 
\frac{r^{p-2}}{(c_{1}r^{p}+1)^{2}} \ .
\end{equation*}
Near the symmetry axis, $r=0$, this function is expanded as
\begin{equation*}
g_{rr} = \frac{c_{1}p^{2}(1-K\kappa)}{2\Lambda} 
r^{p-2} + O(r^{2p-2}) \ .
\end{equation*}
If we introduce a proper radial coordinate $\tilde{X}$ given by
\begin{align}
    \tilde{X} = \sqrt{\frac{2c_{1} (1-K\kappa)}{\Lambda}} 
    r^{p/2} \ ,
\end{align}
we can express the two-metric induced on the $r\phi$-section near the axis $r=0$ in terms of $\tilde{X}$ as
\begin{equation*}
ds_{2}^{2} \approx
d\tilde{X}^{2} + \frac{p^{2}}{4} \tilde{X}^{2} d\phi^{2} \ .
\end{equation*}
Thus, to avoid a conical singularity at $r=0$, which is not desirable, we should impose $p=2$ for the asymptotically de Sitter case. One can easily check that this choice of $p$ gives no angle defect at spatial infinity. 
Now, the equation for $V(r)$ becomes
\begin{align}
V^{\prime \prime} 
+ \bigg(\frac{1}{r} + \frac{2 \big(c_{1}r^{2}-1\big)}{r \big(c_{1} r^{2}+1\big)}
\bigg)V^{\prime}
+ 32 K\kappa c_{1}^{2} \bigg( \sqrt{\frac{K\kappa \lambda} {2\Lambda} } + \lambda\bigg)
\frac{r^{2}}{\big( c_{1 }r^{2}+1 \big)^{4}}  = 0 \ , \label{EqV2} 
\end{align}
whose solution is found to be
\begin{align}
V= - \frac{v_{0}}{\big(1+c_{1}r^{2} \big)^{2}  } - \frac{v_{1}}{1+c_{1}r^{2}}  +v_{2} \ , \label{Vsol3}
\end{align}
where $v_{1}$ and $v_{2}$ are integration constants, and we have defined
\begin{align}
v_{0}= \frac{2(1-K\kappa)}{\Lambda} \ .
\end{align}
In order to avoid infinite dragging of the angular velocity of a co-rotating observer, we set $v_{2}=0$. Also, we can avoid the conical singularity along the symmetry axis by setting $v_{1}=-v_{0}$. Thus, we have
\begin{align*}
V= \frac{2(1-K\kappa)}{\Lambda}\frac{c_{1} r^{2}}{(c_{1}r^{2} +1)^2} \ .
\end{align*}

Remark that the angular velocity of a co-rotating observer is independent of $r$ and is given by
\begin{align}
\omega \equiv - \frac{g_{u\phi}}{g_{\phi\phi}} =
- \dot{G} \ ,
\end{align}
at spatial infinity. Also, the Maxwell equation in Eq. \eqref{maxeq} becomes
\begin{align}
\frac{d^{2}P}{dr^{2}} + \frac{1}{r} \frac{dP}{dr} 
- \frac{p_{1} (c_{1}r^{2}-1)^{2}}{(c_{1}r^{2}+1)^{4}} P = 0 \ ,
\end{align}
where 
\begin{align}
p_{1} = \frac{8 c_1 (1-K\kappa) (2-K\kappa)}{\kappa 
   \Lambda } \ .
\end{align}
Let us introduce a new variable given by
\begin{align}
y= 1- \frac{2}{c_{1}r^{2}+1} \ ,
\end{align}
by which the domain is mapped from $[0, \infty)$ to $[-1,1)$. Correspondingly, the differential equation for $P$ is transformed into
\begin{align}
(1-y^{2})\frac{d^{2}P}{dy^{2}} - 2y \frac{dP}{dy} - c^2 y^{2}P = 0 \ ,
\end{align}
where 
\begin{align}
c^2 =  \frac{2 (1-K\kappa) (2-K\kappa)}{\kappa  \Lambda }\ .
\end{align}
This is the angular part of the zeroth-order spheroidal wave equation since $|y| \leq 1$.  The solution to this equation is given as a series expansion of the Legendre polynomials $P_{k}(y)$:
\begin{align}
P(y) = \sum_{k=0}^{\infty} d_{k} P_{k}(y) \ ,
\end{align} 
where the coefficients $d_{k}$ satisfy the recurrence relation
\begin{align}
c^{2} \alpha_{k-2} d_{k-2} + \left[ k(k+1) + c^{2} \beta_{k}\right] d_{k} + c^{2} \gamma_{k+2} d_{k+2} = 0 \ .
\label{recurrence}
\end{align} 
Here we have defined $d_{-2}=d_{-1}=0$, and $\alpha$, $\beta$, and $\gamma$ are given by
\begin{align}
\alpha_{k} = \frac{(k+1)(k+2)}{(2k+1)(2k+3)} \ , \quad
\beta_{k} = \frac{2k^{2}+2k-1}{(2k-1)(2k+3)} \ , \quad
\gamma_{k} = \frac{k(k-1)}{(2k-1)(2k+1)} \ .
\end{align}
The recurrence relation is of the stride-2 type, and the odd terms are completely decoupled from the even terms. Since our equation is a linear 2nd order ODE, there exist two linearly independent solutions in some neighborhood of a regular point. Thanks to the Birkhoff-Trjitzinsky theorem together with the Pincherle theorem \cite{Gautschi1967,Leaver1985,Benda:2025tni}, we can find a set of two linearly independent solutions, one of which dominates the other: 
\begin{align}
d_{k} = C_{1} f_{k} + C_{2} g_{k} \ , 
\end{align}
for some constants $C_{1}$ and $C_{2}$, and 
\begin{align}
\lim_{k \rightarrow \infty} \frac{f_{k}}{g_{k}} = 0 \ .
\end{align}
Here, $f_{k}$ and $g_{k}$ are called the minimal and dominant solutions, respectively. The exclusion of the dominant solution can be realized by a matching condition of a continued fraction of the ratios of the coefficients \cite{Gautschi1967}. Let us consider the ratio $r_{k} \equiv d_{k}/d_{k-2}$. It follows from Eq. \eqref{recurrence} that the ratio $r_{k}$ obeys the following equation:
\begin{align}
r_{k} = - \frac{c^{2} \alpha_{k-2}}{k(k+1) + c^{2} \beta_{k}
+ c^{2} \gamma_{k+2}r_{k+2}} \ .
\label{ratio}
\end{align}
There are two ways to obtain $r_{2}=d_{2}/d_{0}$. The first way is to read it off from Eq. \eqref{recurrence} directly:
\begin{align}
r_{2} = - \frac{\beta_{0}}{\gamma_{2}} = - \frac{5}{2} \ .
\end{align}
The other way is to compute it by the continued fraction from Eq. \eqref{ratio}. The Pincherle's Theorem states that the dominant solution is excluded when these two values coincide:
\begin{align}
- \frac{5}{2} = \frac{-c^{2}\alpha_{0}}{[6+c^{2}\beta_{2}] - \frac{c^{4} \alpha_{2} \gamma_{4}}{(20+c^{2}\beta_{4}) - \frac{c^{4}\alpha_{4} \gamma_{6}}{(42+c^{2}\beta_{6})} - \cdots}
} \ .
\end{align}
The remaining task is to show that the minimal solution converges at $y=\pm1$. Equivalently, we have to show that
\begin{align}
\sum_{k} |d_{k}| < \infty \ . 
\end{align}
For a sufficiently large $k \gg 1$, the recurrence relation is cast into the following form:
\begin{align}
\frac{c^{2}}{4} r_{k+2} + \frac{c^{2}}{4 r_{k}}  \approx -k^{2}  \ .
\end{align}
Thus, the first or the second term on LHS must cancel $-k^{2}$ on RHS. We have two cases. When the first term cancels the right hand side, we have
\begin{align}
\frac{c^{2}}{4} r_{k+2}  \approx -k^{2} \quad \Longrightarrow \quad    
r_{k+2} \approx - \frac{4k^{2}}{c^{2}} \quad \Longrightarrow \quad  
\frac{d_{k+2}}{d_{k}} \approx - \frac{4k^{2}}{c^{2}} \ .
\end{align} 
It gives a super-exponentially growing mode, which corresponds to the dominant solution. On the other hand, when the second term cancels the right hand side, we have
\begin{align}
\frac{c^{2}}{4 r_{k}} \approx -k^{2} \quad \Longrightarrow \quad    
r_{k} \approx - \frac{c^{2}}{4k^{2}} \quad \Longrightarrow \quad  
\frac{d_{k+2}}{d_{k}} \approx - \frac{c^{2}}{4k^{2}} \ .
\end{align} 
It gives a super-exponentially decaying mode corresponding to the minimal solution. The d'Alembert's ratio test tells us that the series in this case converges since
\begin{align}
\lim_{k \rightarrow \infty} \bigg\vert \frac{d_{k+2}}{d_{k}} \bigg\vert = 0 \ .
\end{align}
The regularity condition of the Maxwell field at $r=0$ requires the following condition:
\begin{align}
    \frac{d P}{dy} \bigg\vert_{y=-1} 
    = \sum_{k=0}^{\infty}  \frac{d P_{k}}{dy}  \bigg\vert_{y=-1} 
    = \sum_{k=0}^{\infty} \frac{(-1)^{k} \, k(k+1)}{2} d_{k} =0 \ . 
\end{align}
The series on the right hand side is convergent since $d_{k}$ chosen above by the matching condition of the continued fraction is super-exponentially decaying. This condition restricts the initial values of the stride-2 three-term recurrence relation.

Now, we are left with only one equation:
\begin{align}
    M''+\frac{M'}{r}-\frac{1}{2}
   \kappa  (P')^2-\frac{4 c_1 (2- \kappa  K) (1-\kappa  K) \left(c_1 r^2-1\right)^2}{\Lambda 
   \left(c_1r^2+1\right)^4}P^2 + \frac{8 c_1
   (1-\kappa  K) \left(c_1 r^2
   \left(c_1r^2-4\right)+1\right)}{\Lambda 
   \left(c_1 r^2+1\right)^4} = 0 \ .
   \label{EqM2}
\end{align}

\subsection{An exact solution without the electromagnetic coupling}

As in the asymptotically flat case analyzed in the previous section, we can explore the scenario where the electromagnetic coupling is turned off. Remarkably, even in the presence of a cosmological constant, the Einstein-Skyrme system admits an exact solution when we set $P=0$. Under this assumption, the exact solution of Eq. \eqref{EqM2} is found to be
\begin{align} \label{Msol2}
M = M_{0} -\frac{\alpha r^{2}}{4 \left(c_1 r^2+1\right)^{2} }  \ ,
\end{align}
where the parameter $\alpha$ is defined in terms of the cosmological constant and the coupling constants as
\begin{align}
\alpha = \frac{8 c_1  (1- K\kappa) }{\Lambda} \ .
\end{align}
Here, we have chosen the integration constants to ensure that the metric function $g_{uu}$ remains regular along the symmetry axis, that is $M^{\prime}(0)=0$.

It follows that this solution has the global minimum at the radial distance
\begin{align*}
    r_{\min}= 1/\sqrt{c_{1}} \ ,
\end{align*}
with the corresponding minimum value of the metric function given by
\begin{align}
M_{\textrm{min}} = M(r_{\textrm{min}}) = M_{0} - \frac{\alpha}{16c_{1}} \ .
\end{align}
Consequently, to prevent the formation of zeros that would define an ergoregion, the function $M$ can be made strictly positive definite everywhere. This regular behavior is achieved simply by choosing the constants such that the following inequality holds:
\begin{align}
M_{0} > \frac{\alpha}{16c_{1}} \ .
\end{align}

\section{Angular momentum of the solutions}

As explained in the preliminaries, because the function $R_2=\dot{G} \, V$ depends non-trivially on $r$, it cannot be gauged away, in our case not even locally. The presence of this irremovable term is responsible for frame-dragging, which implies that the solution is spinning. The angular momentum can then be calculated following an approach analogous to that of Ref. \cite{PoSte}.

\subsection{General setup}

In the presence of a cosmological constant $\Lambda$, the $u\phi$-component of the Einstein field equations reads
\begin{align}
G_{u\phi} + \Lambda g_{u\phi} = \kappa T_{u\phi} \ ,
\end{align}
where $g_{u\phi} = R_{2}$, and the $u\phi$-component of the Einstein tensor for our metric ansatz is given by
\begin{align}
G_{u\phi} = - \frac{1}{2}e^{-R_{1}} \bigg[ R_{2}^{\prime \prime} 
- \frac{1}{r}R_{2}^{\prime} - R_{2} R_{1}^{\prime \prime} - \bigg(R_{2}^{\prime}
+ \frac{1}{r}R_{2}\bigg)R_{1}^{\prime}\bigg] \ .
\end{align}
From the energy-momentum tensor perspective, the net angular momentum density on the hypersurface $u = \textrm{const.}$ is obtained by integrating $T_{u\phi}$ over the entire two-dimensional transverse space:
\begin{align}
\mathcal{J}(\infty) = \int_{0}^{2\pi} d\phi \int_{0}^{\infty} dr \sqrt{\bar{g}} \, T_{u\phi}
= \frac{1}{\kappa} \int_{0}^{2\pi} d\phi \int_{0}^{\infty} dr \sqrt{\bar{g}} \, \big( G_{u\phi} + \Lambda g_{u\phi} \big) \ ,
\end{align}
where the transverse volume element is $\sqrt{\bar{g}} = r e^{R_{1}}$.

First, let us evaluate the Einstein tensor contribution. Using integration by parts, the integrand, $\sqrt{\bar{g}} \, G_{u\phi}$, can be written as an exact total derivative:
\begin{align}
\frac{1}{\kappa} \int_{0}^{2\pi} d\phi \int_{0}^{r} d\tilde{r} \sqrt{\bar{g}} \, G_{u\phi}
=\frac{2\pi}{\kappa} R_{2} + \frac{\pi}{\kappa} r R_{1}^{\prime} R_{2} - \frac{\pi}{\kappa} r R_{2}^{\prime} \ ,
\end{align}
where we have used the regularity conditions at the symmetry axis, $R_{2}(u,0)=0$, $R_{1}^{\prime}(0)=0$, and $\lim_{r \to 0} r R_{2}^{\prime}(u,r) = 0$. Combining this boundary contribution with the cosmological constant term yields the general expression for the total angular momentum density at spatial infinity:
\begin{align}
\mathcal{J}(\infty) = \frac{\pi}{\kappa} \lim_{r \to \infty} \Big[ 2 R_{2} + r R_{1}^{\prime} R_{2} - r R_{2}^{\prime} \Big] 
+ \frac{2\pi \Lambda}{\kappa} \int_{0}^{\infty} dr \, r e^{R_{1}} R_{2} \ .
\label{eq:master_J}
\end{align}

\subsection{Asymptotically flat spacetime}

In the asymptotically flat sector ($\Lambda = 0$, $\lambda = 0$), the second term in Eq.~\eqref{eq:master_J} vanishes identically. For $K\kappa\, p< 1$, the metric function $R_{2}$ is given by
\begin{align}
R_{2} = 2 K \kappa \, \dot{G} \, r^{2} \, {}_{2}F_{1} \big(1+2K\kappa, 2/p;
1+2/p; -c_{1}r^{p}\big) \ .
\end{align}
At spatial infinity ($r \to \infty$), the asymptotic behaviors of the metric functions are
\begin{align}
R_{1}^{\prime} (\infty) \approx - \frac{2K\kappa\, p}{r} \ , \qquad
R_{2}(u,\infty) = V_{\infty} \, \dot{G} \ , \qquad
\lim_{r \to \infty} r \, R_{2}^{\prime} = 0 \ ,
\end{align}
where the asymptotic constant $V_{\infty}$ is
\begin{align}
V_{\infty} = \frac{\Gamma\big(1+\frac{2}{p}\big)\Gamma\big(1+2K\kappa 
- \frac{2}{p}\big)}{c_1^{2/p} \Gamma(2K\kappa)} \ .
\end{align}
Substituting these asymptotic relations into Eq.~\eqref{eq:master_J}, the net angular momentum density at spatial infinity becomes
\begin{align}
\mathcal{J} (\infty)
= \frac{2\pi}{\kappa} \big(1 - K\kappa p \big) V_{\infty}\, \dot{G} \ ,
\end{align}
which is finite and non-vanishing for $K\kappa\, p < 1$.

\subsection{Asymptotically de Sitter spacetime}

In the asymptotically de Sitter sector ($\Lambda > 0$, $\lambda > 0$, $p=2$), the metric functions $e^{R_{1}}$ and $R_{2}$ are given by
\begin{align}
e^{R_{1}} = \frac{2c_{1}(1-K\kappa)}{\Lambda (1 + c_{1}r^{2})^{2}} \ , \qquad
R_{1}^{\prime} = - \frac{4c_{1}r}{1+c_{1}r^{2}} \ ,
\end{align}
and
\begin{align}
R_{2} = \dot{G} \, V \ , \qquad \textrm{where} \qquad
V= \frac{2(1-K\kappa)}{\Lambda}\frac{c_{1} r^{2}}{(1 + c_{1}r^{2})^2} \ .
\end{align}
A direct substitution shows that $2 R_{2} + r R_{1}^{\prime} R_{2} - r R_{2}^{\prime} = 0$ identically, which implies that $G_{u\phi} = 0$ everywhere and that the first term in Eq.~\eqref{eq:master_J} vanishes. However, because $\Lambda \neq 0$ and $g_{u\phi} = R_{2} \neq 0$, the energy-momentum tensor component $T_{u\phi}$ is non-vanishing:
\begin{align}
T_{u\phi} = \frac{1}{\kappa} \big( G_{u\phi} + \Lambda g_{u\phi} \big) = \frac{\Lambda}{\kappa} R_{2}
= \frac{2(1-K\kappa)}{\kappa} \frac{c_{1} r^{2}}{(1 + c_{1}r^{2})^{2}} \dot{G} \ .
\end{align}
Consequently, evaluating the integral over the transverse space up to spatial infinity via Eq.~\eqref{eq:master_J} yields
\begin{align}
\mathcal{J}(\infty) &= \int_{0}^{2\pi} d\phi \int_{0}^{\infty} dr \sqrt{\bar{g}} \, T_{u\phi}
= \frac{2\pi \Lambda}{\kappa} \int_{0}^{\infty} dr \, r e^{R_{1}} R_{2} \notag \\
&= \frac{8\pi c_{1}^{2} (1-K\kappa)^{2}}{\kappa \Lambda} \dot{G} \int_{0}^{\infty} \frac{r^{3}}{(1 + c_{1}r^{2})^{4}} \, dr \ .
\end{align}
Using the exact definite integral
\begin{align}
\int_{0}^{\infty} \frac{r^{3}}{(1 + c_{1}r^{2})^{4}} \, dr = \frac{1}{12 c_{1}^{2}} \ ,
\end{align}
the scale parameter $c_{1}$ cancels out completely, and we obtain a finite, non-zero total angular momentum density at spatial infinity:
\begin{align}
\mathcal{J}(\infty) = \frac{2\pi (1-K\kappa)^{2}}{3\kappa \Lambda} \dot{G} \ .
\end{align}
Thus, from the perspective of the energy-momentum tensor $T_{u\phi}$, both the asymptotically flat and asymptotically de Sitter spinning mesonic tubes carry a well-defined, finite, and non-vanishing angular momentum at spatial infinity.

\section{Energy density and topological charge density}

In this section, we characterize the exact solutions presented above (that is, the solutions without coupling to the electromagnetic field) by analyzing their metric profiles, energy density, total energy and topological charge.

\begin{figure}
    \centering
    \includegraphics[width=0.44\linewidth]{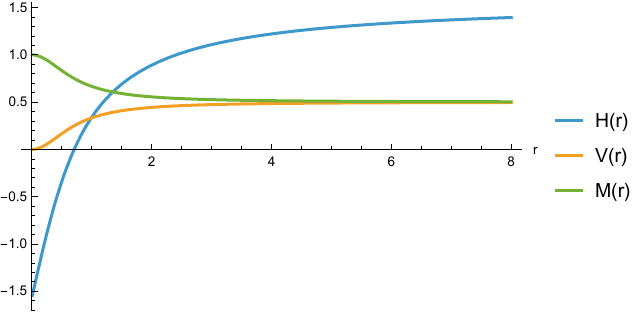} \qquad \includegraphics[width=0.44\linewidth]{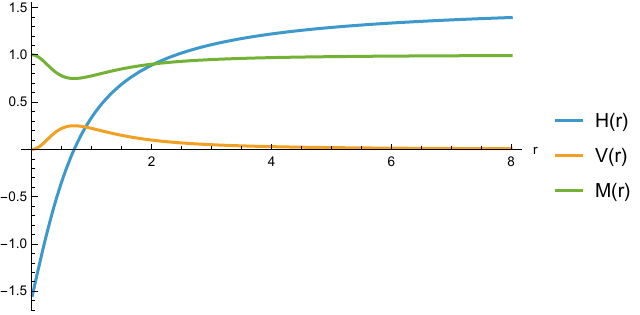}
    \caption{Soliton profile $H(r)$ and metric functions $V(r)$ and $M(r)$ for asymptotically flat (left) and asymptotically de Sitter (right) mesonic tubes.}
    \label{Plots1}
\end{figure}

\begin{figure}
    \centering
    \includegraphics[width=0.36\linewidth]{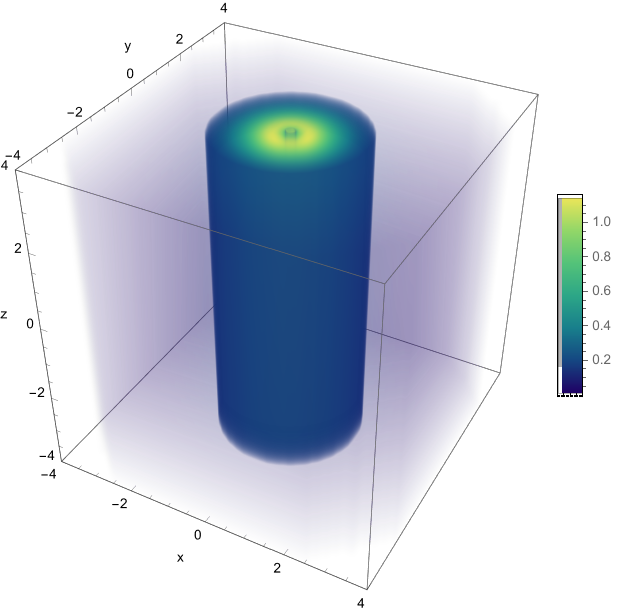}  \qquad \includegraphics[width=0.36\linewidth]{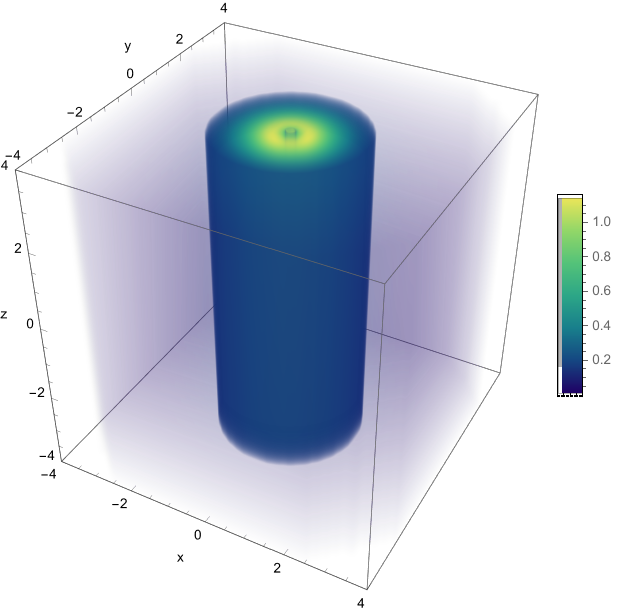}
    \caption{Energy density of asymptotically flat (left) and asymptotically de Sitter (right) mesonic tubes.}
    \label{Plots2}
\end{figure}

First, we note that the asymptotically flat and asymptotically de Sitter configurations exhibit qualitatively similar behavior (differing, of course, in their asymptotic regions). As pointed out in previous sections, both spacetime geometries and matter fields remain regular at the origin, smooth throughout the spatial domain, and well determined by imposing the boundary conditions. The metric functions $V(r)$ and $M(r)$, as well as the matter field profile $H(r)$, are shown in Fig.~\ref{Plots1}.
The plots have been obtained using the exact solutions in Eqs.~\eqref{Hsol}, \eqref{Vsol1}, \eqref{Msol1}, \eqref{Vsol3} and \eqref{Msol2}, which are compatible with the boundary conditions $M(0)=1$, $M'(0)=0$, $V(0)=0$, and $V'(0)=0$. 
For both cases, we fulfill the condition $\kappa=1/(K p)$, and we have fixed $K=0.1$, $p=2$, and $c_1=2$. In the asymptotically de Sitter case, we have used $\Lambda=1$ and $\lambda=0.25$ to satisfy the constraint in Eq.~\eqref{constraintcouplings}.
The energy density, $T^u{}_u$, for the general ansatz described in Eqs.~\eqref{metric} and \eqref{U2} takes the form
\begin{equation}  \label{generalED}
    T^u{}_u = - \frac{K}{8r^2} e^{-2R_1}\biggl( p^2e^{R_1} \cos^2(H) + \bigl(4r^2e^{R_1} + p^2 \lambda \cos^2(H)\bigr) (H')^2 \biggr) \ .
\end{equation}
Figure~\ref{Plots2} shows the three-dimensional energy density profile, derived from the analytical expressions in Eq.~\eqref{generalED}, illustrating a hollow tubular structure where the maximum value of the energy density is located at a finite radius from the origin and decays asymptotically at large distances. This shell-like distribution is qualitatively reminiscent of vortex-like or cosmic string configurations, where the core remains empty and the energy is localized around the symmetry axis.
From Eqs.~\eqref{energy} and \eqref{generalED}, we can compute the energy of these configurations. For the asymptotically flat solutions, the energy per unit length is given by 
\begin{equation}
\mathcal{E}_{\textrm{AF}} = 2\pi K p \ , 
\end{equation}
while for the asymptotically de Sitter configuration, the energy reads  
\begin{equation}
\mathcal{E}_{dS} = 2 \pi K p  \biggl( 1 + \sqrt{\frac{\Lambda \lambda }{2K \kappa}} \biggr) \ . 
\end{equation}
Note that the difference between $\mathcal{E}_{dS}$ and $\mathcal{E}_{\textrm{AF}}$ arises precisely from the positive contribution of the cosmological constant $\Lambda$ modulated by the Skyrme coupling $\lambda$. 
Evaluating at $p=2$, which is the case of interest for the exact solutions, we obtain $\mathcal{E}_{\textrm{AF}}= 4 \pi K$ and $\mathcal{E}_{dS} = 4 \pi K \biggl( 1 + \sqrt{\frac{\Lambda \lambda }{2K \kappa}}\biggr)$.

Furthermore, we examine the topological properties of the Skyrme matter field. The total topological charge $B$ is defined by
\begin{equation}
B = \int d^3x \, \mathcal{B}^0 \ , 
\end{equation}
where the topological charge density $\mathcal{B}^0$ is given by
\begin{equation}
\mathcal{B}^0 = \frac{1}{24\pi^2}\epsilon^{0ijk}\mathrm{Tr}\left[(U^{\dagger}\partial_i U)(U^{\dagger}\partial_j U)(U^{\dagger}\partial_k U)\right] \ .
\end{equation}
Interestingly, while the topological charge density $\mathcal{B}^0(r,\phi)$ is non-vanishing, the total charge itself vanishes ($B=0$). This cancelation arises from the spatial symmetry of the field configuration, where regions with positive and negative topological densities balance each other upon integration, reflecting a topologically non-trivial local field structure with zero net global charge. 

\section{Conclusions}

In this manuscript, we have constructed novel self-gravitating configurations within the $SU(2)$ Einstein non-linear sigma and Einstein-Skyrme models that asymptotically behave as gyratons while maintaining regular curvature invariants. These analytic solutions, which can also include minimal coupling to electrodynamics, describe spinning pulsed mesonic tubes. They represent generalized gyratons, as their sourcing energy-momentum tensor includes components beyond those of a standard null fluid. Crucially, the gyratonic nature of these tubes manifests in their non-trivial angular momentum, driven by an off-diagonal frame-dragging effect that cannot be gauged away.
The matter distribution supporting these geometries is entirely singularity-free, with its maximum energy density concentrated in a hollow tubular region around the propagation axis. By analyzing the $SU(2)$ NLSM minimally coupled to General Relativity and the Maxwell field, we first obtained asymptotically flat configurations. A remarkable feature of these solutions is that they can exhibit compact support in the light-cone time coordinate, ensuring that both their total energy and angular momentum remain finite.
Subsequently, we demonstrated that incorporating the Skyrme term forces the spacetime to be asymptotically de Sitter in order to preserve the exact resolvability of the field equations. Interestingly, while the total topological charge of the system strictly vanishes, the local topological charge density remains non-zero.
As a future perspective, recent developments in Refs. \cite{BPS1,BPS2,BPS3,BPS4} could allow the present approach to be generalized to include additional matter fields. In particular, it would be highly compelling to incorporate complex scalar fields describing superfluid condensates, where the physical vorticity naturally connects to the gyratonic nature of the configurations. This intriguing issue is currently under investigation.

\acknowledgments

The authors are grateful to Fabrizio Canfora for his collaboration and valuable contributions in the early stages of this work. A.G. has been funded by FONDECYT Regular No. 1240247,  M. L. has been funded by FONDECYT Iniciaci\'on No.  11241079. A. V. has been funded by FONDECYT Iniciaci\'on No.  11261883. S. H. O. is supported 
by the National Research Foundation of Korea funded by the Ministry 
of Science and Technology (Grant RS-2025-16071544).


\newpage

\appendix

\section{Formal series solution for the Maxwell field} \label{AppB}

For $p \neq 2$, the Maxwell field equation is given by
\begin{align}
    \frac{d^{2}P}{dz^{2}} + \frac{1-2z}{z(1-z)} \frac{dP}{dz} - \eta \frac{z^{\zeta}(2z-1)^{2}}{z^{2}(1-z)^{2}}P = 0 \ , 
\end{align}
where we have defined the constant $\eta$ and the non-integer power $\zeta$ as
\begin{align}
    \eta = \frac{4K}{p^{2}c_{1}^{2/p}} \ , \qquad \zeta = \frac{2}{p} \ .
\end{align}
The boundaries $z=0$ and $z=1$ correspond to the origin and spatial infinity, respectively.
To accommodate the non-integer $\zeta$, we assume a double Frobenius series solution:
\begin{align}
    P(z) = \sum_{m=0}^{\infty} \sum_{n=0}^{\infty} a_{n,m} z^{n + m\zeta + s} \ .
\end{align}
Substituting this ansatz into the differential equation and aligning the powers of $z$ yields the recurrence relation for the combined exponent $\lambda_{n,m}(s) = n + m\zeta + s$:
\begin{align}
    [\lambda_{n,m}(s)]^2 a_{n,m}(s) 
    =& \, [\lambda_{n,m}(s)-1][2\lambda_{n,m}(s)-1] a_{n-1,m}(s) 
    - [\lambda_{n,m}(s)-2][\lambda_{n,m}(s)-1] a_{n-2,m}(s) \notag \\
    &+ \eta \Big( a_{n,m-1}(s) - 4a_{n-1,m-1}(s) + 4a_{n-2,m-1}(s) \Big) \ ,
\end{align}
where $a_{n,m} = 0$ for any $n < 0$ or $m < 0$. 
The indicial equation at the origin ($n=0, m=0$) yields a repeated root $s = 0$. 
Because the physical Maxwell field must be completely regular at the origin $z=0$, the logarithmically divergent second solution is discarded. 
Evaluating the base layer ($m=0$) with $s=0$ forces $a_{n,0} = 0$ for all $n \ge 1$. 
Thus, the physically regular formal series solution reduces to:
\begin{align}
    P(z) = a_{0,0} \left( 1 + \sum_{m=1}^{\infty} \sum_{n=0}^{\infty} a_{n,m} z^{n + m\zeta} \right) \ .
\end{align}
All higher-order coefficients $a_{n,m}$ are completely determined by the cross-coupling of the fractional layers ($m \ge 1$) in the recurrence relation.




\begin{thebibliography}{999}

\bibitem{AiSe} P. C. Aichelburg and R.U. Sexl, Gen. Rel. Grav. 2 (1971) 303.

\bibitem{EaGi}D.M. Eardley and S.B. Giddings, Phys. Rev. D 66, 044011 (2002).

\bibitem{YoNa1}H. Yoshino, and Y. Nambu Phys. Rev. D 66 065004 (2002).

\bibitem{YoNa2}H. Yoshino, and Y. Nambu Phys. Rev. D 67 024009 (2003).

\bibitem{YoRi}H. Yoshino, and V. S. Rychkov Phys. Rev. D 71 104028 (2005).

\bibitem{YoZe}H. Yoshino, A. Zelnikov, V. Frolov, Phys. Rev. D75 124005 (2007).

\bibitem{FePe}V. Ferrari and P. Pendenza, GRG 22 1105 (1990).

\bibitem{SaLo}N. Sanchez and C. Lousto, Nucl. Phys. B383 377 (1992).

\bibitem{Bo}W.B. Bonnor, Int. J. Theor. Phys. 3, 257 (1970).

\bibitem{FroFu}V.P. Frolov and D.V. Fursaev, Phys. Rev. D 71,104034  (2005).

\bibitem{FroIsZe}V. Frolov, W. Israel, A. Zelnikov, Phys. Rev. D\textbf{72},  084031 (2005).

\bibitem{FroZe1} V.~P.~Frolov and A.~Zelnikov,
Phys. Rev. D \textbf{72}, 104005 (2005).

\bibitem{FroZe2} V.~P.~Frolov and A.~Zelnikov,
Class. Quant. Grav. \textbf{23}, 2119-2128 (2006).

\bibitem{PoSte}J. Podolsky, R. Steinbauer, R. Svarc, Phys. Rev. D 90, 044050 (2014).

\bibitem{PoSva}J. Podolsky, R. Svarc, Phys. Rev D 99 044004, (2019).

\bibitem{KrPoZeKa1}P. Krtous, J. Podolsky, A. Zelnikov, H. Kadlecova, Phys. Rev. D 86, 044039 (2012).

\bibitem{KrPoZeKa2}H. Kadlecova, A. Zelnikov,  P. Krtous, J. Podolsky ,Phys.Rev.D 80 (2009) 024004.

\bibitem{PoSvaMa}J. Podolsky, R. Svarc, H. Maeda, Class. Quant. Grav. 36 (2019) 1, 015009.

\bibitem{FroLi}V.P. Frolov and F.L. Lin,  Phys. Rev. D73 104028 (2006).

\bibitem{CaKleZo}M. Caldarelli, D. Klemm, E. Zorzan, Class. Quant. Grav. 24 (2007), 1341-1357.

\bibitem{Jetz} P. Jetzer, Phys. Rept. 220 (1992), 163-227.

\bibitem{Nair} V.P. Nair, "Quantum field theory a modern perspective", Springer (2005).

\bibitem{ManSu} N. Manton, P. Sutcliffe, "Topological solitons", Cambridge University Press (2007)

\bibitem{Canf1} F. Canfora, Phys. Rev. D88, 065028 (2013).

\bibitem{CanfMaed} F. Canfora, H. Maeda, Phys. Rev. D87 094049 (2013).

\bibitem{ACGO}
M.~Astorino, F.~Canfora, A.~Giacomini and M.~Ortaggio,
Phys. Lett. B \textbf{776}, 236-241 (2018).

\bibitem{CDGP} F. Canfora, N. Dimakis, A. Giacomini, A. Paliathanasis, Phys. Rev. D99 (2019), 4, 044035.

\bibitem{GO} A. Giacomini, M. Ortaggio, JHEP 09 (2019) 090.

\bibitem{CGLOV} F.~Canfora, A.~Giacomini, M.~Lagos, S.~H.~Oh and A.~Vera,
Eur. Phys. J. C \textbf{81}, no.1, 55 (2021).

\bibitem{Bonga:2022nwe}
B.~Bonga and G.~Dotti,
Phys. Rev. D \textbf{105}, no.4, 044049 (2022).

\bibitem{Ayon-Beato:2015eca}
E.~Ayon-Beato, F.~Canfora and J.~Zanelli,
Phys. Lett. B \textbf{752}, 201-205 (2016).

\bibitem{Ayon-Beato:2019tvu}
E.~Ay\'on-Beato, F.~Canfora, M.~Lagos, J.~Oliva and A.~Vera,
Eur. Phys. J. C \textbf{80}, no.5, 384 (2020).

\bibitem{Astorino:2018dtr}
M.~Astorino, F.~Canfora, M.~Lagos and A.~Vera,
Phys. Rev. D \textbf{97}, no.12, 124032 (2018).

\bibitem{Henriquez-Baez:2022ubu}
C.~Henr\'\i{}quez-B\'aez, M.~Lagos and A.~Vera,
Phys. Rev. D \textbf{106}, no.6, 064027 (2022).

\bibitem{Concha:2023qcs}
P.~Concha, C.~Henr\'\i{}quez-Baez, E.~Rodr\'\i{}guez and A.~Vera,
Phys. Rev. D \textbf{107}, no.12, 124043 (2023).

\bibitem{Henriquez-Baez:2024rjb}
C.~Henr{\'\i}quez-Baez, M.~Lagos, E.~Rodr{\'\i}guez and A.~Vera,
Phys. Rev. D \textbf{111}, no.10, 104016 (2025).

\bibitem{Vera:2025qqz}
A.~Vera,
JHEP \textbf{10}, 119 (2025).

\bibitem{Canfora:2026wfi}
F.~Canfora, A.~Neira and S.~H.~Oh,
Eur. Phys. J. C \textbf{86}, no.6, 613 (2026).

\bibitem{Canfora:2026dym}
F.~Canfora, C.~Corral and B.~Diez,
Phys. Rev. D \textbf{114}, no.4, 044084 (2026).

\bibitem{Canfora:2026col}
F.~Canfora, A.~Gomberoff, C.~Henr{\'\i}quez-Baez and A.~Vera,
Phys. Rev. D \textbf{113}, no.8, 084068 (2026).

\bibitem{BPS1}
F.~Canfora,
JHEP 11 (2023) 007.

\bibitem{BPS2}
F.~Canfora, M.~Lagos and A.~Vera,
JHEP \textbf{10}, 224 (2024),

\bibitem{BPS3}
F.~Canfora,  P.~Pais,
Nucl. Phys. B 1017 (2025) 116955.

\bibitem{BPS4}
F.~Canfora, P.~Pais,
Eur. Phys. J. C 85 (2025) 8, 884.

\bibitem{BPS5}
S. L. Cacciatori, F. Canfora, E. Delgado, F. Muscolino, L. Rosa,
Phys. Rev. D 113 (2026) 10, 105025.

\bibitem{BPS6}
F.~Canfora, M.~Garrido, M.~Mannarelli and A.~Neira,
Phys. Rev. D \textbf{114}, no.7, 074011 (2026).

\bibitem{SU(N)1}
P.~D.~Alvarez, S.~L.~Cacciatori, F.~Canfora and B.~L.~Cerchiai,
Phys. Rev. D \textbf{101}, no.12, 125011 (2020).

\bibitem{SU(N)2}
S.~L.~Cacciatori, F.~Canfora, M.~Lagos, F.~Muscolino and A.~Vera,
JHEP \textbf{12}, 150 (2021).

\bibitem{SU(N)3}
S.~L.~Cacciatori, F.~Canfora, M.~Lagos, F.~Muscolino and A.~Vera,
Nucl. Phys. B \textbf{976}, 115693 (2022).

\bibitem{Heun}A. Ronveaux, {\it Heun’s Differential Equations}, Oxford University Press (1995).

\bibitem{Gautschi1967}
W. Gautschi, 
SIAM Rev. \textbf{9} (1967), no. 1, 24--82.

\bibitem{Leaver1985}
E. W. Leaver, 
Proc. R. Soc. Lond. A \textbf{402} (1985), no. 1823, 285--314.

\bibitem{Benda:2025tni}
K.~Benda and J.~Matyjasek,
Phys. Rev. D \textbf{111}, no.12, 124010 (2025).


\end{thebibliography}
\end{document}